# Using a small number of devices to experimentally estimate the packet delivery ratio on a lorawan network with a large number of end devices

Maxim S. Ilderyakov
*Saint Petersburg State University
of Aerospace Instrumentation*
Saint Petersburg, Russia
maxim.ilderyakov@gmail.com

Nikita V. Stepanov
*Saint Petersburg State University
of Aerospace Instrumentation*
Saint Petersburg, Russia
nstepanov@k36.org

*Abstract*—In this paper, the proposed methodology for conducting experiments on a small number of devices for evaluating systems with a large number of devices. A software package for assessing the performance of LPWAN (Long Range Wide Area Networks, LoRaWAN) networks is presented. An example of the operation of this software system as applied to the LoRaWAN network for estimating the probability of message delivery from a number of devices to a base station when implementing this technique is shown. This technique can be used to evaluate the performance of other LPWAN networks.

## I. INTRODUCTION

In recent years, interest to LoRaWAN technology is growing. The technology has been many advantages before other similarity technologies. It is a low energy expenses, high noise immunity, opportunity of using different spreading factors(SF) for separation channels. LoRaWAN is not full investigated, for example you can't find information about interaction of a lot of devices and how the devices can interfere to each other. This work sets the task to answer for the question, how a lot of devices will lead itself with interfere to each other. This task is very important for evaluating network performance.

Key parameter to give assessment of effective system is probability successful delivery of packets from devices to base station(BS) . Parameters will be fix on LoRaWAN-server. We will give assessment of probability successful delivery in real experiments.

During the experiments to BS are sends packets from fix quantity of devices. Information from BS is need to use into the enter of the program complex. Experiment is held with matching of Algorithm of program work and with output information we can give assessment of probability successful delivery. In real experiments we can't use a lot of device . In this paper is described  the melodic allowing using to replace 4 devices  in experiment on 1000 devices in real system.

## II. RATIONALE FOR USE APPROACH TO THE ORGANIZATION OF THE EXPERIMENT

By simulating a large number of message overlaps in a real system, we obtain the ratio of a single device in an experimental system to the number of devices in a real system, about 4: 1000.If all devices use one SF and last selected so that in the absence of overlaps is messages successfully delivered, we can calculate it with formulas (1) and (2)

$$\Pr\{T = 2tmes.\} = e^{-\lambda N2t}, \quad (1)$$

T - overlap of message when message is not recognized.

In situation when messages can't be recognized with full message overlay,  probability successful delivery can be find with the theoretic formula(2)

$$\Pr\{T = 2tmes\} = e^{-\lambda Nt}. \quad (2)$$

From formulas (1) and (2) we can see that packet delivery ratio is affects on production  $e^{-\lambda Nt}$ . In experiment we increase lenth of message and reduce period.

For needed provision of intensive we enter into consideration the next surveys: in real systems can used interval of transmission messages from one device $T_p$=10 min. So, we have intensive of transmission message $\lambda_p$=1/600. In bounds of explore system number of real devices  must be $N_p$.=10000, from here we have next intensive:

$$t_p \cdot N_p \cdot \lambda_p = 687,$$

where $t_p$  is duration of short packet.

This parameters of real system we can watch find in BS [5]and end devices.

In experiment we need reduce time of message retransmissions and to increase length of message. Given the limitations of configuration firmware and characteristics ED the interval of transmission messages from one device $T_e=7$ sec.[6] From here we have intensity of transmission messages $\lambda_e=1/7$, ED number $N_e=41$. Intensity in experimental explore equals

$$T_e \cdot N_e \cdot \lambda_e \approx t_p \cdot N_p \cdot \lambda_p,$$

where $t_e$ – duration of long packet.

So, we imitation a large number of messages overlaps in real system, then we have next ratio: four messages in experimental system for thousand messages in real system.

### III. DESCRIPTION OF THE SOFTWARE- HARDWARE COMPLEX

In order to assess the effective of system work we was developed the program complex, it must simplify hard work of get and processing data. The complex manages the course of the experiment, get information from LoRaWAN-server and calculates the probability of delivery. In this complex built in algorithm of devices off.

To program input you must set table with description of experiment. It have list of devices what will using in the experiment. The program do brute force of devices and offer to operator on each iteration turn on the new device. If operator made a mistake and don't turn one device, we can see this error on output file. When all devices are turn on, operator set the command to start experiment. Experiment time is input parameter and operator can change it.

On the second computer must be start the server. Server has access to log files of BS, where information of all devices is written.

When experiment time will be end program start to get and processing data. With help of server API program with websocket protocol connect with server and send json-request for each device. Server answer have counter of packet. The numeration of packet goes in order and we can calculate how many packet was not sent on BS.

After parsing server answer and data processing, information is write in output file. The format of the file have next view – all strings has three values – device id, number of delivered packages and number of sent packages. This format is comfortable for the next work.

When experiment will be end, the program will offer to operator turn off devices by some algorithm. This algorithm will consider later.

The program complex can divide into two parts: control of the experiment and data processing.

### IV. CONTROL OF EXPERIMENT

Control of experiment have three stages: stage of turn on devices, stage of waiting when experiment time will be end and stage of turn off devices. All stages will be discussed in detail later.

### V. ALGORITHM OF TURN ON DEVICES

To start the program you must set two tables in xls format. The first table have id of devices what will be used in the experiment, the second table have two values in all strings : device EUI and device ID. ID is easier to work with it than EUI and this table intended to ease the operator's work. EUI is used to send request to server.

After start program will brute force devices. Program get device id and matches it with EUI, this values will saved in memory as matrix. The matrix has size 2xN, where N is number of uses devices in the experiment. The matrix will be used to make the transition from ID to EUI.

The devices will start in the order they are located in the table. If device was not turn on this information will be write in output file. This information will be used later in algorithm of turn off devices.

### VI. WAITING OF EXPERIMENT END

When matrix is ready, program save time of experiment start and find time of experiment end. Then program wait the end of experiment. When time is over program do bust of matrix strings. The one iteration create the request for one device. All requests have next important values: device EUI, time of start and time of end experiment. This values allow take accurate selection from the time. Server's anwers go to the parsing. When parsing is end results write to output file.

To secondary output will be write timestamp information about all packets. This file is auxiliary to look the course of the experiment.

### VII. ALGORITHM OF TURN OFF DEVICES

To the forming queue of device turn off, all devices will be splitting into three sets: set with high priority, set with middle priority and set with low priority. Next will be reviewed algorithm of turn off devices.

Step 1.Turn off devices with high priority. In this moment some part of out file is finished to used, this part used to forming first queue of devices to shutdown . In this queue input devices which was answer while the experiment was going on. The program asks to disable devices from this queue.

If after shutdown some device another device starts responding, this device put into queue with middle priority.

Step 2.Turn off devices with middle priority. When first queue will be empty , starts shutdown devices from second queue.

Step 3.Turn off devices with low priority. Last to shutdown will devices from third queue.

### VIII. DATA PROCESSING

When needed data saved in memory data processing is start data processing. The program goes through all values of packets counter, if some values will be missed, program calculate how many packet was missed.

To calculate how many packets was delivery program get from server's answer number of all packets from experiment time. This value is a number of delivery packets.

End view of output file: information about the experiment, then information about devices which was answer after turn off some device.

## IX. Features of construction program complex

Architecture of program complex is server-client. The complex is client of LoRaWAN-server. Server get json request. When server connection established will start authorization process, if this process is succeed program take request to server. Server's answer will go to parsing. This answer have json format. In the string some values are extras, it will be cut off. After parsing program calculate packets and write it to file.

## X. Experiment's results

In work[7] is scenario with 1000 devices in identical radio conditions send short messages on BS with SF7. With some assumption was shown what part of devices can be change SF from 7 to 8 and it will increase the likelihood of delivery messages. On Fig. 1 you can see dependence of the upper and lower bounds of the probability of success from changed devices number with SF8. In the left point on the horizontal axis value 0 corresponds to 8835 with SF7 and 0 device in real system.

In accordance with the above method for experiment we was take 36 devices. It was configuration to simulate 8835 devices of real system. Number of devices with SF7 and SF8 you can see in the table 1

TABLE I    NUMBER OF DEVICES

|  | Number of devices | | | | |
|---|---|---|---|---|---|
| Number of devices with SF7 in real system | 36 | 31 | 22 | 14 | 7 |
| Number of devices with SF8 in real system | 0 | 1227 | 1718 | 3436 | 5399 |
| Number of devices with SF7 in experiment | 8835 | 7608 | 5399 | 3436 | 1718 |
| Number of devices with SF8 in experiment | 0 | 5 | 7 | 14 | 22 |

The results of experiment you can see on Fig. 1. The dotted line indicated the upper bound of the probability of success, solid line indicated the lower bound of the probability of success, circles indicated the experiment result.

Result of experiment shows that idea from work[7] can up the probability of success . The optimal combination of devices with SF7 and SF8 with fewer of them is changed from optimal value in work[7]. This is due to the fact that in [7] a number of assumptions are used that are not valid for a real system.

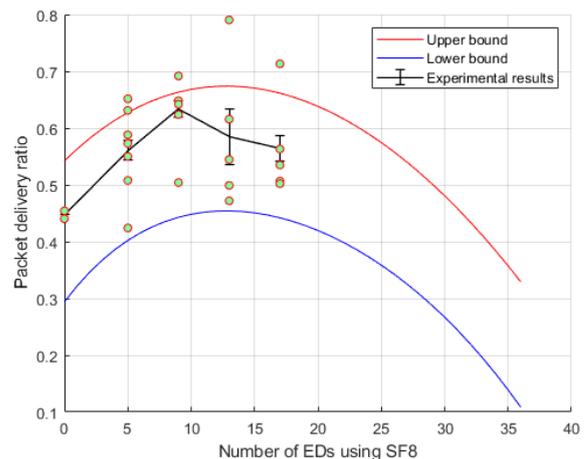

Fig. 1 Influence 36 ED on PDR to network with two SF(SF7 and SF8).

## Conclusion

The program complex can greatly facilitate the work of the experimenter and give assessment of effectively LPWAN network. Output data has convenient format for next data processing.

Effectively of this program follows from matches the program results with theoretic calculations.

## Acknowledgment

The work created in bounds of initiative the science project № 8.8540.2017 / 8.9 «Development algorithms of sends data in IoT systems with accounting to complexity of devices». Authors thanks organization «AURORA Mobile Technologies»( Ltd «IoT Laboratory ») for equipment and special thanks to general director of business development Gusev Oleg Valerievich for and research assistance.